\documentclass[rnote]{aa}  

\usepackage{graphicx}
\usepackage{color}
\usepackage{txfonts}
\begin{document}

   \title{The Nature of Faint Fuzzies from the Kinematics of NGC\,1023}


   \author{Chies-Santos, A. L.,
          \inst{1}
          Cortesi, A.,
    	 \inst{1}
          Fantin,  D. S. M.,
          \inst{1}
          Merrifield, M. R.,
          \inst{1}
        Bamford, S.,
          \inst{1}
          and 
           Serra, P.
         \inst{2}
         \inst{3}
          }
   \institute{University of Nottingham, School of Physics and Astronomy, University Park, NG7 2RD Nottingham,UK\\ 
   \and {Netherlands Institute for Radio Astronomy (ASTRON), Postbus 2, NL-7990 AA Dwingeloo, the Netherlands}
   \and{CSIRO Astronomy and Space Science, Australia Telescope National Facility, PO Box 76, Epping, NSW 1710, Australia}
              }
 \offprints{ana.chies\_santos@nottingham.ac.uk}            
              
   \date{Received 29 August 2013/Accepted 16 October 2013}

 
  \abstract{Faint fuzzies are metal-rich apparently-old star clusters with unusually large radii (7-15 pc), found mostly in S0 galaxies, whose source remain obscure.  To identify their origins, we compare planetary nebulae and neutral hydrogen with faint fuzzy positions and line-of-sight velocities in NGC\,1023.  In this way, we rule out scenarios in which these objects are associated with an on-going merger or with a spheroid population in NGC\,1023.  Their kinematics are indistinguishable from the stellar disk population in this galaxy, and we conclude that faint fuzzies are most likely just remnant open clusters.  Their observed association with S0s then simply reflects the difficulty of identifying such objects in later-type disk galaxies.}

   \keywords{galaxies: elliptical and lenticular, cD, galaxies: star clusters, galaxies: structure}

\titlerunning{The Nature of Faint Fuzzies}
   
\authorrunning{Chies-Santos, Cortesi, Fantin et al.}

   \maketitle
%

\section{Introduction}

Faint fuzzy star clusters (hereafter FFs) were discovered by \cite{LB00} in the discs of the nearby S0 galaxies NGC\,1023 and NGC\,3384. They were reported to be red, old clusters (8-13\,Gyr), with unusually large radii (7-15 pc), and a mass similar to that of halo globular clusters ($10^{5\mbox{-}6}$ M$_\odot$). Subsequent photometric studies suggested the presence of FF-like star clusters in several galaxies in the ACS Virgo Cluster Survey (\citealt{peng06}), in NGC\,5195 (in interaction with M\,51 \citealt{HL06}) and in NGC\,1380 (\citealt{chies07}). In terms of their hosts, it is notable that, in the ACS Virgo Cluster Survey, nine of the galaxies containing possible FFs are S0s, and four are dusty early-type galaxies.

Their large ages imply that they must be long-lived systems akin to globular clusters (hereafter GCs; \citealt{BL02}).
However, there is a great deal of uncertainty in measuring the ages of star clusters. Even in the Milky Way, the exact age difference between different GCs has become clear only recently: \cite{hansen13} have just been able to reach the level of precision necessary to show that the red Galactic GC 47\,Tuc is 2 Gyr younger than the metal-poor NGC\,6397. The situation in extragalactic studies remains far less certain, with conflicting reports of ages for different sub-populations of GCs and the oldest open clusters, leaving a wide range of possibilities to explain the origins of FFs.

The main possibilities are that FFs are residual very old open cluster complexes left over in the disks of these S0s \citep{bruns09}, that they form a new sub-component of the GC system, or that they are associated with merger events, either originating from the merged system [as found in M31 \citep{mackey10}] or forming in the merger itself.  This last possibility would require the apparent ages to be misleading, but the on-going merger of NGC\,1023 with NGC\,1023A \citep{capa86} and the well-documented formation of unusual star clusters in merging systems \citep{ws95} make this scenario worthy of consideration, and render NGC\,1023 a very useful test case.

Clearly the kinematics of FFs should allow us to distinguish between these theories: close relationships with either open or globular clusters should have a clear signature of disk or spheroid kinematics, while a merger origin would follow the distinct kinematics of tidally-stripped streams.  The first clues in this regard came from \citet{BL02}, who found that the kinematics of FFs in NGC\,1023 show significant rotation, which seems to follow the stellar kinematics as found in the absorption-line data of \citet{simien97}.  Unfortunately, however, the FFs are mostly found at larger radii than were reached by these conventional spectra, so this inferrence relied on an extrapolation of the stellar kinematics.  Stellar rotation curves are known to show a variety of behaviours at large radii (\citealt{coccato10}, \citealt{arnold10}, \citealt{foster11}), so the match between FF and stellar kinematics remains uncertain; this uncertainty is unfortunate, as it is quite possible that the FFs display a stronger asymmetric drift than the stellar kinematics -- a smaller net rotation velocity at the same radius -- which would again shift their identification to a more disk-like GC population akin to the metal-rich GCs in the Milky Way \citep{zinn85}.  Such conventional stellar kinematics also lack the subtlety required to identify and trace fainter kinematic features such as the tidal streams resulting from mergers, so one cannot use these data to determine whether the FF kinematics match that of any material being stripped from NGC\,1023A in order to test that scenario.

  \begin{figure*}[ht]
   \centering
\resizebox{0.8\hsize}{!}{\includegraphics[keepaspectratio, angle=0]{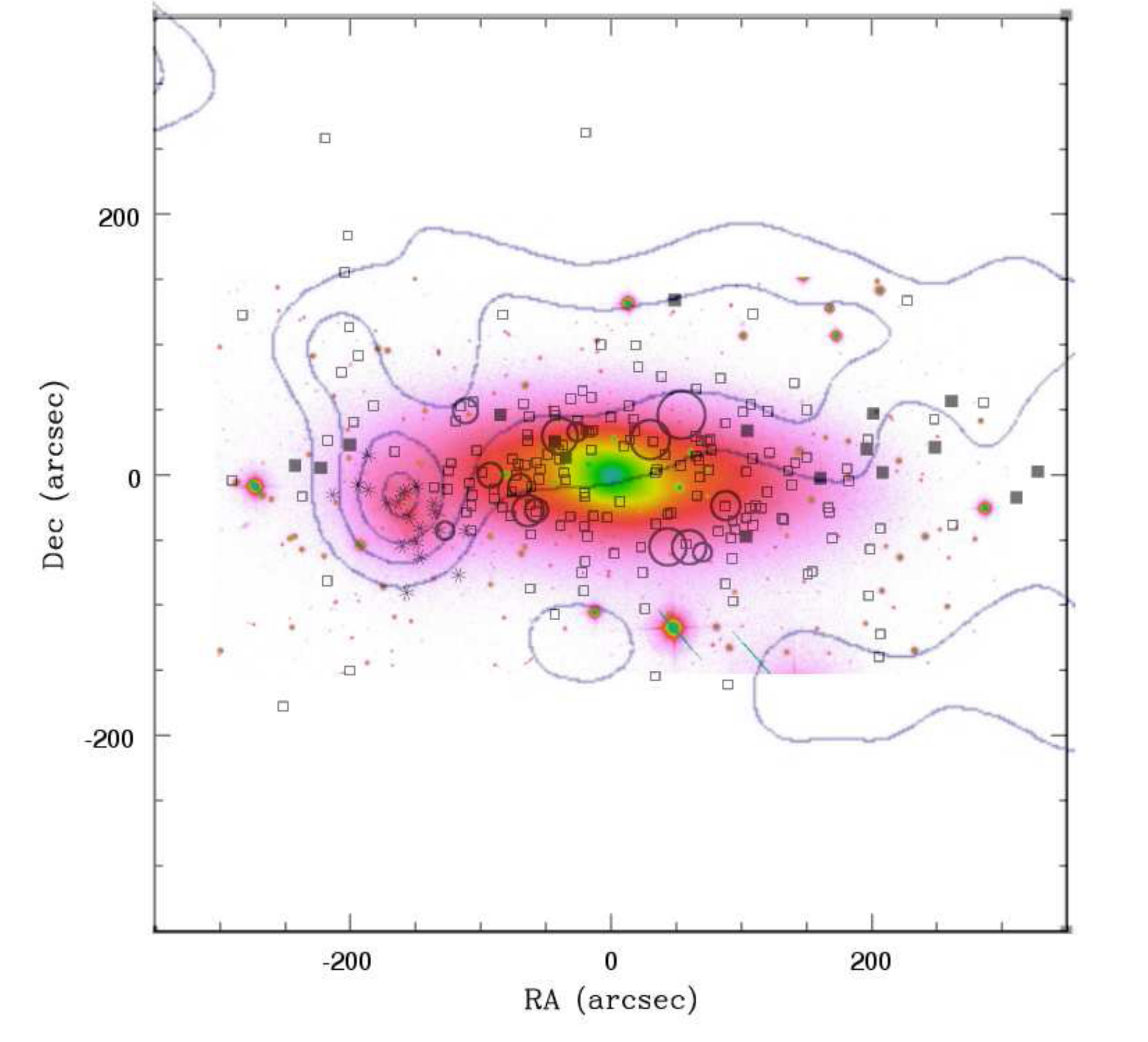}}
\caption{RA and Dec of the PNe (open squares) and the FFs (large open circles, whose radii are proportional to their effective radii) overplotted on a B-band image of the NGC\,1023 system. The companion galaxy NGC\,1023A is evident on the left side of NGC\,1023, as a brighter cloud. The stars over this overdensity indicates the PNe associated to the companion galaxy. Filled squares represent the stream of PNe. The contours represent the {\sc H\,i} distribution (see text for details).}
\label{fig:pos}%
    \end{figure*}

To address these issues, in this Research Note we look at alternate kinematic tracers in NGC\,1023.  Specifically, we use planetary nebulae (PNe) from \citet{noord08} to extend the stellar kinematics to the larger radii occupied by the FFs found by \citet{BL02}.  In this way, we can directly compare their kinematics at similar radii.  Further, we can use the approach of \citet{cortesi11} to separate disk and spheroid stellar kinematics, to see which sub-component the FFs more closely follow.  Finally, these data also allow us to trace subtler kinematic features such as the stellar stream arising from the on-going merger with NGC\,1023A, allowing a direct test of any role it may play in the formation of FFs.  To further explore the possibility that FFs may form in such mergers, we also compare to the {\sc H\,i} data from \citet{morganti06} for this system, to see if there is a supply of gas in the merger from which tidal clusters could be forming.  In Section \ref{sec:position}, we describe the spatial distribution of these data, while in Section \ref{sec:velocity}, we explore their kinematics to infer the nature of FFs.  In Section \ref{sec:conclusion} we discuss possible formation scenarios for FFs in light of these results.

\section{Spatial Distribution}\label{sec:position}

Figure \ref{fig:pos} summarizes the locations of the tracers adopted in this study of NGC~1023.  The underlying B-band image, from \citet{noord08}, has been false-coloured to show both the over-all structure of the galaxy and the companion NGC\,1023A.  Also shown are the locations of the PNe from \citet{noord08} and contours showing the highest column densities of {\sc H\,i} data from \cite{morganti06}.  It is immediately apparent that NGC\,1023A contains significant {\sc H\,i}, and that there appears to be a tidal tail of gas being stripped from this system, stretching right across the north side of the galaxy, leaving open the possibility that the FFs could be associated with such a merger stream.

Indeed, it is notable that a great fraction of the FFs lie close to the gas distribution.
They do appear somewhat offset from the maximum of the {\sc H\,i} emission, but so are the PNe associated with this stream (see Section~\ref{sec:velocity}), shown as filled squares in Fig.~\ref{fig:pos} -- if they do trace the same feature, it would appear that collisional processes are separating the gas from the stellar components.  However, the presence of some FFs to the south of the galaxy perhaps fits better with the ring picture considered by \citet{bbl05}.  Some care is needed in assuming a ring-like morphology, though, as the absence of FFs at small radii could just reflect the difficulty of detecting such objects against the bright galaxy light in this region; the similar apparent lack of PNe at small radii arises from exactly this bias.
  
Clearly, the main message here is that even with these more extensive observations, there is not enough information in the spatial data to identify the FFs with any particular component of NGC\,1023, so we therefore now turn to the extra discrimination provided by the kinematics.

  \begin{figure*}[ht]
   \centering
\resizebox{0.7\hsize}{!}{\includegraphics[keepaspectratio, angle=0]{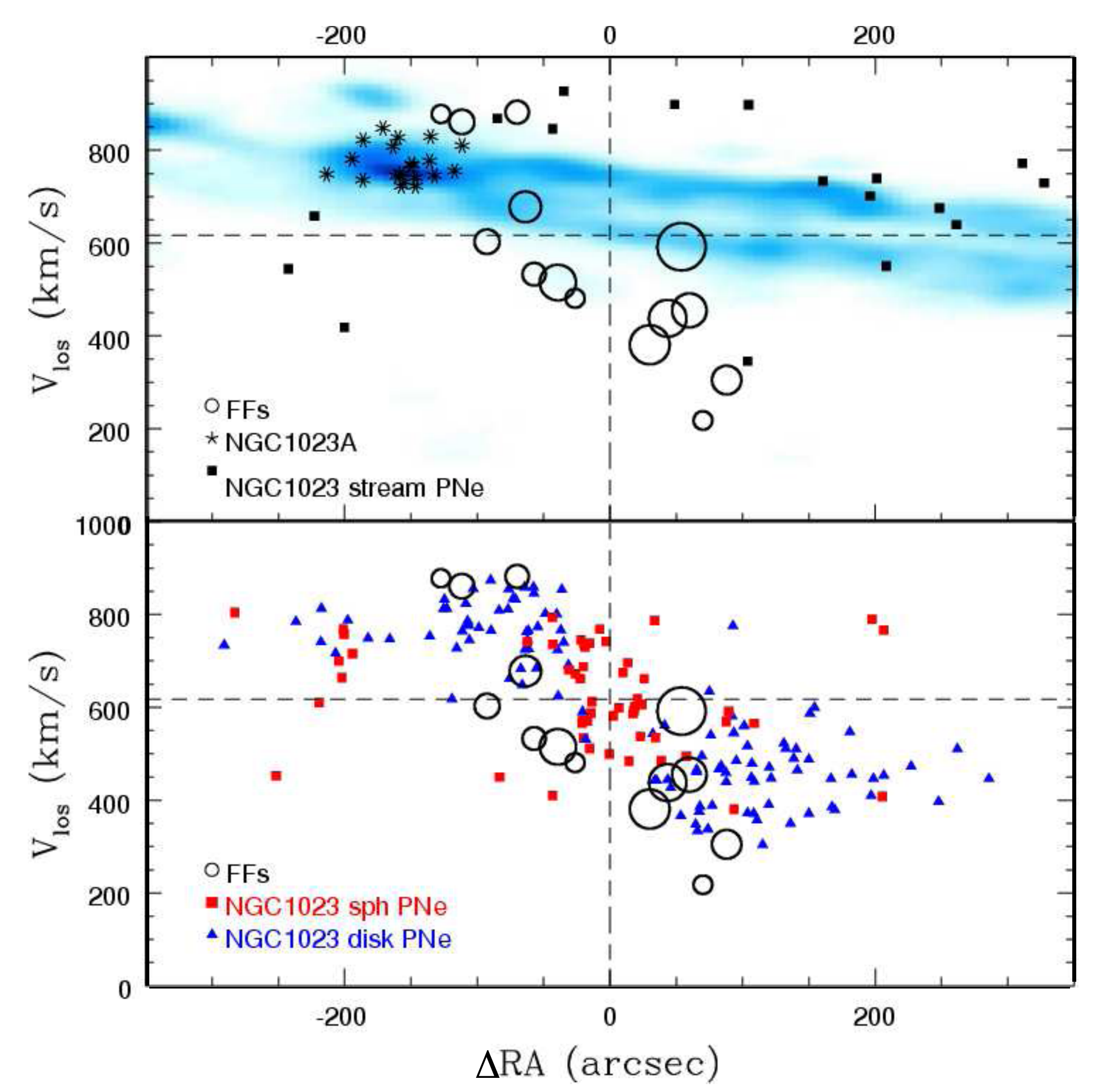}}
\caption{\textit{Top panel:} A ($\Delta RA$, v$_{los}$) diagram  presenting the FFs (red circles), the PNe associated with the companion NGC\,1023A (stars) and the ones associated to the stream (squares) firstly reported by \cite{cortesi11}. The {\sc H\,i} map (\citealt{morganti06}), obtained by collapsing a masked version of the {\sc H\,i} cube along the declination axis is the blue shaded distribution.
\textit{Bottom panel:} The same as the \textit{top panel}, but showing the FFs (red circles), the spheroid (green squares) and disc (blue triangles) PNe belonging to NGC\,1023. Note that the dimension of the symbol of the FFs is proportional to their $R_{eff}$, but not scaled to the figure. For NGC\,1023, 1$\arcsec=50$\,pc.}
  \label{fig:vlos}%
    \end{figure*}

\section{Tracer Kinematics}\label{sec:velocity}
Figure~\ref{fig:vlos} shows the line-of-sight velocities of the various tracers as a function of right ascension (corresponding to position along the major axis given the orientation of this galaxy).  The upper panel shows the {\sc H\,i} data along with the FFs, and the PNe picked out by \citet{noord08} as likely being members of NGC\,1023A as well as those identified by \citet{cortesi11} as not forming part of the regular disk and bulge of NGC\,1023 on the basis of their kinematics and positions.  This plot immediately clarifies the situation: the {\sc H\,i} stream passes right through the velocity of NGC\,1023A, confirming the association of the two, and the PNe with peculiar kinematics also largely follow the stream kinematics, with the same kind of modest offset effects seen in the spatial data.  However, the FFs, although matching the position and velocity of NGC\,1023A, soon part company with the stream, having completely the opposite sign of velocity relative to NGC\,1023 on the far side of the system.  The FFs are clearly not associated with any tidal stripping of NGC\,1023A.

The lower panel of Fig.~\ref{fig:vlos} compares the FF kinematics to those of the PNe that \citet{cortesi11} found did fit with a simple disk-plus-spheroid model of NGC\,1023, with different symbols identifying with which component the PNe are most likely associated.  Again, the picture is fairly clear: the FF kinematics follow those of the disk PNe rather closely. 
One caveat is that the PNe catalogue is incomplete close to the centre of the galaxy, due to the difficulties of detecting such faint objects against the bright galaxy continuum. Moreover the number of PNe associated with the disk component decreases toward the central regions. Consequently, it is not trivial to understand the FFs kinematics in the very centre of the galaxy. Nevertheless, the upper envelope of velocities for the FFs, on both sides of the galaxy, match those found in disk PNe, implying a similar low level of asymmetric drift in both.  Interestingly, though, the FFs do not display the characteristic kinematics that one might expect for a cold ring of clusters, where they would lie on a simple linear feature in this diagram; they seem to be spread over  range of intrinsic radii.  Nonetheless, the FFs unequivocally form a cold disk population.

\section{Discussion}\label{sec:conclusion}

In this paper, we have been able to confirm the impression that FFs form a population that is indistinguishable from the stellar disk population in NGC\,1023, and are not consistent with an association with either the spheroid or the merging companion.  They appear to be simply old open clusters.  

Such a simple identification then raises the question of why they only seem to be found in S0 galaxies.  The apparent ring-like morphology led \citet{bbl05} to suggest that they were formed in such arrangements as part of the transformation of spirals into S0s.  However, the kinematics of the FFs seems more consistent with the general stellar disk population rather than a discrete ring, suggesting that the apparent morphology may simply arise from the difficulty of detecting FFs against the bright stellar background at smaller radii.  Such biases also suggest a simpler explanation as to why these objects are only found in S0s: these faint objects will always be easier to pick out against the smooth background of an S0 than the messier light distribution in the disk of a spiral galaxy.  If very careful studies of spirals ultimately reveal that they contain the progenitors FFs as well, then S0 galaxies may offer the cleanest laboratory in which to study the properties of the longest-lived open clusters formed in spiral galaxies.

\begin{acknowledgements} 
We thank Søren Larsen for providing tables with sizes for NGC\,1023 FFs and Jean Brodie, Hanni Lux and Evelyn Johnston for useful discussions. We have made use of the WSRT on the Web Archive. The Westerbork Synthesis Radio Telescope is operated by the Netherlands Institute for Radio Astronomy ASTRON, with support of NWO. PS is a NWO/Veni fellow.
\end{acknowledgements}

\end{document}